\documentclass[10pt]{iopart}
\usepackage{epsfig}

\begin{document} 
%\wideabs{
\title{A technique to directly excite Luttinger liquid collective
   modes in carbon nanotubes at GHz frequencies}

%\author{P.J. Burke\\
%Integrated Nanosystems Research Facility\\
%Department of Electrical and Computer Engineering\\
%University of California, Irvine, 92697}

\author{P.J. Burke}
\address{Integrated Nanosystems Research Facility\\
Department of Electrical and Computer Engineering\\
University of California, Irvine, 92697}

%\date{\today} 
%\ead{pburke@uci.edu}
%\maketitle

\begin{abstract}
We present a technique to directly excite Luttinger liquid 
collective modes in carbon nanotubes at GHz frequencies.  
By modeling the nanotube as a
nano-transmission line with distributed kinetic and magnetic
inductance as well as distributed quantum and electrostatic
capacitance, we calculate the complex, frequency dependent impedance
for a variety of measurement geometries. Exciting voltage waves on the
nano-transmission line is equivalent to directly exciting the yet-to-be
observed one dimensional plasmons, the low energy excitation of a
Luttinger liquid. Our technique has already been applied to 2d
plasmons and should work well for 1d plasmons.
Tubes of length 100 microns must be grown for
GHz resonance frequencies. Ohmic contact is not necessary with our
technique; capacitive contacts can work.
\end{abstract}

\pacs{\footnote{This work has been submitted to the IEEE for possible
publication.}PACS: 61.46,73.63.Fg,85.85.+j,71.10.Pm}
%}
%\twocolumn

\section{Introduction}

One of the most fundamental unsolved questions in modern condensed
matter physics is: What is the ground state of a set interacting
electrons, and what are the low-lying excitations? By far the most 
successful theoretical treatment of interactions is Landau's theory of 
Fermi liquids, which posits that the low-lying excitations of a Fermi 
liquid are not in fact electrons, but "quasiparticles" which, to good 
approximation, are non-interacting. The reason that the quasiparticles 
can be treated as non-interacting is that the inverse quantum lifetime 
of a quasiparticle is generally less than its
energy, so that the concept of an independent quasiparticle is well
defined. Landau's Fermi liquid theory has served physicists well in
two and three dimensions for many decades. Unfortunately, it has long
been known that Landau's Fermi liquid theory breaks down in
one-dimensional systems\cite{Fisher}, such as those formed in single walled carbon
nanotubes (SWNTs)\cite{Dresselhaus}.

To deal with this problem, Tomonaga\cite{Tomonaga}, and later
Luttinger\cite{Luttinger},
described a simplified model for interacting electrons in one dimension, which 
was exactly solvable.  The method used was that of
bosonization\cite{Gogolin,Mahan}. 
The boson variables describe collective excitations in the
electron gas, called {\it 1d plasmons}. Later, Haldane~\cite{Haldane} argued that the
bosonization description was generically valid for the low energy
excitations of a 1d system of interacting electrons, coining the term
the ``Luttinger liquid''. In this model, the creation of an electron
is equivalent to exciting an infinite number of 1d plasmons.  Much
theoretical work\cite{Fisher} has gone into calculating the experimental
consequences of the non-Fermi liquid behavior of 1d systems. The main
experimental consequences calculated and observed\cite{BockrathNature} to date are the
power-law dependence of conductivity on temperature and the power-law
dependence of tunneling current on
bias voltage, when tunneling from a 3d macroscopic lead into the 1d
system. The power-law exponent is generally 
characterized by a dimensionless parameter ``g''.  For non-interacting
electrons, g=1, while for interacting electrons, $g<1$. To date, 
the experimental evidence for the theory that the low-lying excitations 
of interacting electrons in 1d are collective plasmon oscillations, while
significant, is somewhat indirect.

It is the purpose of this paper to describe a technique that can be
used to directly excite the 1d plasmons using a microwave signal
generator. (Similar proposals have appeared in the literature 
already.\cite{Ponomarenko,Sablikov,Blanter})
This technique was recently applied to measure collective
oscillations (plasmons) in a two-dimensional electron gas, including
measurements of the 2d plasmon velocity, as well as the temperature
and disorder dependent damping\cite{BurkeAPL}. 
Our goal in this paper is to describe a technique
to extend these measurements to one-dimensional systems, and to
discuss a method to directly measure the 1d plasmon velocity, 
and hence ``g'' in a Luttinger liquid.
In order to discuss this technique, one of our goals in this paper is
to provide an effective circuit model for the effective electrical 
(dc to GHz to THz) properties 1d interacting electron systems. 
While we restrict our
attention to metallic SWNTs, the general approach can be used to describe
semiconducting carbon nanotubes, multi-walled carbon nanotubes, quantum
wires in GaAs heterostructures\cite{Auslaender}, and any other system
of one-dimensional interacting electrons.

In our recent 2d plasmon work, we suggested a transmission-line
effective circuit model to relate our electrical impedance measurements to the
properties of the 2d plasmon collective 
excitation\cite{BurkeAPL,BurkeUnpublished,PeraltaAPL,PeraltaThesis}.
There, we measured the kinetic inductance of a two-dimensional
electron gas, as well as its distributed distributed 
electrostatic capacitance to
a metallic ``gate'' by directly exciting it with a microwave voltage.
The distributed capacitance and inductance form a transmission line,
which is an electrical engineer's view of a 2d plasmon.

Since then, the transmission-line description has
been discussed in the context of both
single-walled\cite{BockrathThesis} and
multi-walled\cite{Tarkiainen,Sonin} carbon nanotubes. In
reference~\cite{BockrathThesis},
 by considering the Lagrangian of a one-dimensional
electron gas (1DEG), an expression for the {\it distributed} 
quantum capacitance (which was not
important in our 2d experiments) as well as the {\it distributed} 
kinetic inductance of 
a SWNT is derived. 
(In references~\cite{Cuniberti1996,Cuniberti1998}, the concept of a
lumped (as opposed to distributed) quantum capacitance and
quantum/kinetic inductance is introduced.)
In reference~\cite{Tarkiainen,Sonin} the
tunnel conductance at high voltages is related to electrical
parameters (the characteristic impedance) of the
transmission line in a multi-walled nanotube. In both of these
discussions, the distributed inductance and capacitance per unit length
form a transmission line, which is again an electrical engineer's
description of a 1d plasmon. It is the goal of this manuscript to
describe how we can excite 1d plasmons directly with a microwave
voltage, calculate the expected results for a variety of possible
measurement geometries (including capacitive as well as tunneling
electrical contacts), and discuss how our technique can be used to
directly measure the Luttinger liquid parameter ``g''.

We proceed as follows: First, we re-derive the results of
reference~\cite{BockrathThesis} for
a spinless 1d quantum wire, by calculating the kinetic inductance,
electrostatic capacitance, and quantum capacitance per unit length. We
extend the results of reference~\cite{BockrathThesis} by considering the magnetic inductance
per unit length, as well as the characteristic impedance. We then
proceed to discuss spin-1/2 electrons in a SWNT, and derive four
coupled equations for the voltages on each of the four quantum
channels in a SWNT, following reference~\cite{Sonin}. We diagonalize these equations of
motion and solve for the spin/charge modes. These results are not
meant to be rigorous many-body calculations, but a way to translate
theoretical ideas about interacting electrons
in 1d into measurable predictions. For more rigorous discussions, the
reader is referred to 
references~\cite{Safi,Cuniberti1996,Cuniberti1998,Ponomarenko,Sablikov,Blanter,SablikovJLTP}.

In the second section of the paper, we proceed to discuss our
technique to directly excite these 1d plasmons by setting up
standing-wave resonances in SWNTs of finite length, as we did in the
2d plasmon case. We
calculate explicitly measurable electronic
properties of 1d plasmons that are amenable to the measurement
technique we developed for 2d plasmons, including the nanotube
dynamical impedance (real and imaginary) as a function of frequency,
as well as the 1d plasmon damping, wave velocity, and Luttinger ``g''
factor. We discuss what experimental parameters are needed to
perform our experiment, and also how the low (sub-GHz) frequency properties of
nanotubes may be used to give some insight into the 1d plasmon.
Finally, we discuss possible practical consequences\cite{Dyakonov2001} 
of the results in
nanotube electronic and micro/nano-mechanical high-frequency circuits.
Our measurement technique could provide direct evidence for
collective mode behavior of interacting electrons in one dimension,
the ``Luttinger liquid''.

In what follows we will define each symbol uniquely to avoid confusion
and ambiguity.  If there seems
to be a ``plethora'' of symbols, it is due in part to the author, and
in part to the complexity of mother nature.

\section{Circuit Model for spinless electrons in a one-channel quantum wire}

\begin{figure}
\epsfig{file=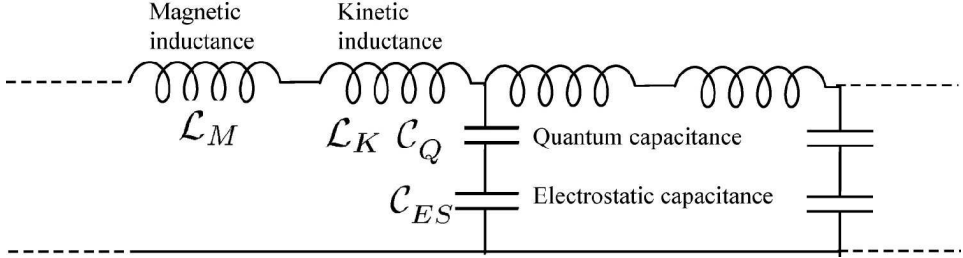}
\caption{Circuit diagram for 1d system of spinless electrons. Symbols are defined per unit length.}
\label{fig:spinlesscircuit}
\end{figure}

The dc circuit model for a one-channel quantum wire of non-interacting
electrons is well known from the Landauer-B\"uttiker formalism 
of conduction in quantum systems.  The dc conductance is simply
given by $e^2/h$. If the spin degree of freedom is accounted for,
there are two ``channels'' in a quantum wire: spin up and spin down,
both in parallel. We postpone our discussion of spin until the next
section, and assume for the moment the electrons are spinless.
At ac, the circuit model is not well established experimentally.  However, theoretically
it is believed to be equivalent to a transmission line, with a
distributed ``quantum'' capacitance and kinetic inductance per unit
length.  It is generally believed\cite{Fisher} that the effect of
electron-electron interactions can be included in the transmission
line circuit analogy as an electrostatic capacitance.  Furthermore,
there will also be a magnetic inductance.

\begin{figure}
\epsfig{file=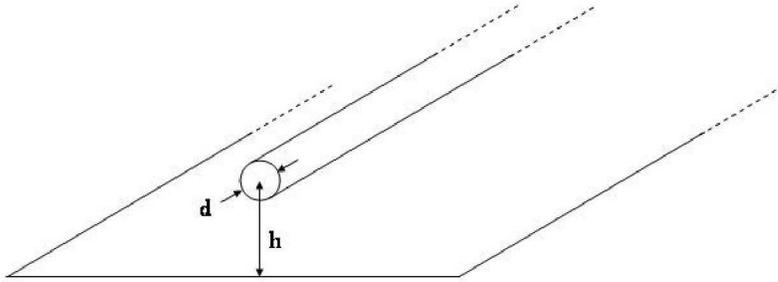}
\caption{Geometry of nanotube in presence of a ground plane.}
\label{fig:geometry}
\end{figure}

The effective circuit diagram we are proposing is shown in
figure~\ref{fig:spinlesscircuit}. Below, we will discuss each of the
four contributions to the total circuit, and then discuss some of its
general properties, such as the wave velocity and characteristic
impedance. We will restrict ourselves to the case of a wire over a
``ground plane'' for the sake of simplicity.  If there is no ground
plane, the the parameter ``h'' below (the distance from the wire to
the ground plane) should be replace by the length of the 1d wire itself.
The geometry we consider is shown in figure~\ref{fig:geometry}.

\subsection{Magnetic Inductance}

In the presence of a ground plane, the magnetic inductance per unit
length is given by\cite{Ramo}:
\begin{equation}
\label{eq:lmagnetic}
{\mathcal L}_M = {\mu\over 2 \pi}~cosh^{-1}\biggl({2h\over
  d}\biggr)\approx {\mu \over 2 \pi}~ln\biggl({h\over d}\biggr),
\end{equation}
where d is the nanotube diameter and h is the distance to the ``ground plane.''
The approximation is good to within 1~\% for $h> 2d$.
This is calculated using the standard technique of setting the inductive energy
equal to the stored magnetic energy:
\begin{equation}
{1\over 2}L I^2={1\over 2\mu}\int B(x)^2~d^3x,
\end{equation}
and using the relationship between $I$ and $B$ in the geometry of
interest, in this case a wire on top of a ground plane.
For a typical experimental situation, the nanotube is on top of an
insulating (typically oxide) substrate, with a conducting medium
below. (The finite conductivity of the conducting medium will be
discussed below.) A typical oxide thickness is between 100~$\AA$ and
1~$\mu m$, whereas a typical nanotube radius is 1~nm. Because the
numerical value of ${\mathcal L}_M$ is only logarithmically sensitive
to the ratio of $d/h$, we can estimate it within a factor of three as:

\begin{equation}
\label{eq:lmagneticnumerical}
{\mathcal L}_M \approx 1~pH/\mu m.
\end{equation}

We use $\mu m$ for our length units because modern growth processes
produce nanotubes with lengths of order microns and not (as of yet) meters.

\subsection{Kinetic Inductance}
In order to calculate the kinetic inductance per unit length, we
follow reference~\cite{BockrathThesis} 
and calculate the kinetic energy per unit length and
equate that with the ${1\over 2}LI^2$ energy of the kinetic inductance. The
kinetic energy per unit length in a 1d wire is the sum of the kinetic
energies of the left-movers and right-movers. If there is a net
current in the wire, then there are more left-movers than
right-movers, say.  If the Fermi level of the left-movers is raised by
$e\Delta\mu/2$, and the Fermi-level of the right-movers is decreased
by the same amount, then the current in the 1d wire is
$I=e^2/h\Delta\mu$. The net increase in energy of the system is the
excess number of electrons ($N = e \Delta \mu /2 \delta$) in the left vs right moving states times
the energy added per electron $e \Delta \mu / 2$. Here $\delta$ is the
single particle energy level spacing, which is related to the Fermi
velocity through $\delta = \hbar v_F 2 \pi /L$. Thus the excess kinetic
energy is given by $hI^2/4v_Fe^2$. By equating this energy with the
${1\over 2}LI^2$ energy, we have the following expression for the kinetic
energy per unit length:
\begin{equation}
\label{eq:lkinetic}
{\mathcal L}_K = {h\over 2 e^2 v_F} 
\end{equation}
The Fermi velocity for graphene and also carbon nanotubes is usually
taken as $v_F=8~10^5~m/s$, so that numerically
\begin{equation}
{\mathcal L}_K =  16~nH/\mu m.
\end{equation}
It is interesting to compare the magnitude of the kinetic inductance
to the magnetic inductance. From equations~\ref{eq:lmagnetic} and
\ref{eq:lkinetic}, 
we have
\begin{equation}
{ {{\mathcal L}_M}\over{ {\mathcal L}_K} }=\alpha~{2\over \pi}~{v_F\over
  c}~ln\biggl({h\over d}\biggr)\sim 10^{-4},
\end{equation}
where $\alpha\approx 1/137$ is the fine structure constant.  
Thus, in 1d systems, the kinetic inductance will always dominate.
This is an important point for engineering nano-electronics: In
engineering macroscopic circuits, long thin wires are usually
considered to have relatively large (magnetic) inductances. In the
case of nano-wires, the magnetic inductance is not that important; it
is the kinetic inductance that dominates.

\subsection{Electrostatic capacitance}
The electrostatic capacitance between a wire and
a ground plane as shown in figure~\ref{fig:geometry} is given by\cite{Ramo}
\begin{equation}
\label{eq:celectrostatic}
{\mathcal C}_E = {2 \pi \epsilon \over
  cosh^{-1}\bigl(2h/d\bigr)}\approx {2\pi\epsilon \over ln(h/d)},
\end{equation}
where again the approximation is good to within 1~\% for $h> 2d$.
This can be approximated numerically as
\begin{equation}
\label{eq:cesnumerical}
{\mathcal C}_E \approx 50~aF/\mu m,
\end{equation}
This is calculated using the standard technique of setting the capacitive energy
equal to the stored electrostatic energy:
\begin{equation}
\label{eq:esenergy}
{Q^2\over 2 C} = {\epsilon\over 2}\int E(x)^2~d^3x,
\end{equation}
and using the relationship between $E$ and $Q$ in the geometry of
interest, in this case a wire on top of a ground plane. The term
``electrostatic'' comes from the approximation that we make in
calculating the capacitance using the above equation~\ref{eq:esenergy},
which is done using the relationship between a static charge and a
static electric field.
However, the electrostatic capacitance can of course be
used when considering time-varying fields, voltages, currents, and
charges, as we will do below.

\subsection{Quantum capacitance}
In a classical electron gas (in a box in 1,2, or 3 dimensions), to add
an extra electron costs no energy. (One can add the electron with any
arbitrary energy to the system.)  In a quantum electron gas (in a box
in 1,2, or 3 dimensions), due to the Pauli exclusion principle it is
not possible to add an electron with energy less than the Fermi energy
$E_F$.  One must add an electron at an available quantum state above $E_F$.
In a 1d system of length L, the spacing between quantum states is
given by:
\begin{equation}
\delta E = {dE\over dk} \delta k = \hbar v_F {2\pi\over L},
\end{equation}
where L is the length of the system, and we have assumed a linear
dispersion curve appropriate for carbon nanotubes.
By equating this energy cost with an effective 
quantum capacitance\cite{BockrathThesis,Tarkiainen,Blanter}
with energy given by
\begin{equation}
{e^2\over C_Q}=\delta E,
\end{equation}
one arrives at the following expression for the (quantum) capacitance
per unit length:
\begin{equation}
{\mathcal C}_Q = {2 e^2\over h v_F}, 
\end{equation}
which comes out to be numerically
\begin{equation}
{\mathcal C}_Q = 100~aF/\mu m.
\end{equation}
The ratio of the electrostatic to the quantum capacitance is then
given by
\begin{equation}
{{\mathcal C}_{ES}\over{\mathcal C}_Q}={2\pi h\over e^2\mu
  v_F}~ln\biggl({h\over d}\biggr)
={1\over \alpha}~{2\over
  \pi}~{v_F\over c}~ln\biggl({h\over d}\biggr)~\sim~1.
\end{equation}
Thus, when considering the capacitive behavior of nano-electronic
circuit elements, both the quantum capacitance and the electrostatic
capacitance must be considered.

\subsection{Wave velocity}

For a distributed inductance and capacitance per unit length, a
technique used by theorists is to write down the Lagrangian (kinetic
minus potential energy), and then to use
the Euler-Lagrange equations to derive an equation of motion which, in
this case, ends up being a wave equation. However, a much simpler
(if somewhat less rigorous)
approach is simply to use a result known by rf engineers for many
decades, namely that the wave velocity of a circuit with distributed
inductance and capacitance is given by:
\begin{equation}
v_{general}=\sqrt{1\over {\mathcal L C}}.
\end{equation}
If we consider only the magnetic inductance (neglecting the kinetic
inductance) and if we also consider only the electrostatic capacitance
(neglecting the quantum capacitance), then the wave velocity would
simply by the speed of light c:
\begin{equation}
v_{free space}=\sqrt{1\over {\mathcal L}_M {\mathcal C}_{ES}}=\sqrt{1\over \mu \epsilon}=c.
\end{equation}
A full solution to the collective mode of a carbon nanotube should
include both the kinetic inductance as well as the magnetic
inductance, which we write as
\begin{equation}
{\mathcal L}_{total}={\mathcal L}_K + {\mathcal L}_M,
\end{equation}
as well as both the quantum capacitance and the electrostatic
capacitance, which we write as
\begin{equation}
{\mathcal C}_{total}^{-1}={\mathcal C}_Q^{-1} + {\mathcal C}_{ES}^{-1}.
\end{equation}
In our recent work\cite{BurkeAPL} 
on a two-dimensional electron gas system (in the presence
of a ground plane), we found that the kinetic inductance dominates
${\mathcal L}_{total}$, and that the geometric capacitance dominates
${\mathcal C}_{total}$, so that the collective mode velocity in 2d is given by:
\begin{equation}
v_{2d} \approx \sqrt{1\over {\mathcal L}_K {\mathcal C}_{ES}}.
\end{equation}
However, as our estimates above show, for a 1d quantum system such as
a nanotube, the quantum capacitance is predicted to dominate ${\mathcal C}_{total}$,
so that in 1d we have the approximation that
\begin{equation}
v_{1d,non-interacting} \approx \sqrt{1\over {\mathcal L}_K {\mathcal C}_Q}=v_F.
\end{equation}
One method of including the effect of 
electron-electron interactions in the context of the above discussion
is simply to include the electrostatic capacitance as well as the
quantum capacitance, so that the wave velocity is not quite exactly
equal to the Fermi velocity:

\begin{equation}
v_{1d,interacting} \approx \sqrt{1\over {\mathcal L}_K {\mathcal C}_{total}} = \sqrt{{1\over {\mathcal L}_K {\mathcal C}_{ES}}+{1\over {\mathcal L}_K {\mathcal C}_Q}}<v_F.
\end{equation}
The ratio of the plasmon velocity in the presence of interactions to
the plasmon velocity in the absence of interactions has a special
significance, and it is given in this simple model by:
\begin{equation}
g_{spinless} \equiv {v_{F}\over v_{1d,interacting}} = \biggl({1+{{\mathcal C}_Q
      \over{\mathcal C}_{ES}}}\biggr)^{-1/2} =
\biggl(1+\alpha~{\pi\over 2}~{c\over v_F}~{1\over ln(h/d)} \biggr)^{-1/2},
\end{equation}
(We use the
subscript spinless to differentiate $g_{spinless}$ from a different g which we
define below.) This immediately suggests a technique to search for
Luttinger liquid behavior in order to measure $g_{spinless}$, namely to measure the
wave velocity. According to these calculations, the measured wave
velocity should differ from the Fermi velocity by a large factor, of order unity.
(If the distance to the ground plane becomes
larger than the tube length such as in some free-standing carbon
nanotubes\cite{Dai}, 
another formula for the capacitance has to be
used, which involves replacing h with the length of the 1d wire.)
Finally, we note that the full solution to the wave velocity is given by
\begin{eqnarray}
v_{1d,interacting} = \sqrt{ 1\over {\mathcal L}_{total} {\mathcal C}_{total}  }
= \sqrt{{1\over  ({\mathcal L}_K + {\mathcal L}_M)}\biggl( {1\over
    {\mathcal C}_Q}+{1\over{\mathcal C}_{ES}} \biggr)            }\nonumber\\
=v_F~\sqrt{ {1+\alpha{\pi\over 2}{c\over
      v_F}ln(d/h)}\over{1+\alpha{2\over \pi}{v_F\over c}ln(h/d)} }.
\end{eqnarray}
With this the g factor should read:
\begin{equation}
\label{eq:gspinless}
g_{spinless} = \biggl({{1+\alpha{\pi\over 2}{c\over
      v_F}ln(d/h)}\over{1+\alpha{2\over \pi}{v_F\over c}ln(h/d)}}\biggr)^{-1/2}
\end{equation}
To our knowledge this full function has not been discussed in the
literature.  We speculate that $g_{spinless}$ should be redefined as in the above
equation~\ref{eq:gspinless} to include this term, which is equivalent to adding the
magnetic energy term to the Hamiltonian. 

The definition of g in a quantum wire when the spin degree of freedom is taken into
account will be discussed in further detail below. For now, we would
like to address the question which naturally arises in the context of
this discussion, how to observe these collective excitations?
One technique, which we propose here, is to measure
the wave velocity in the frequency or time domain.
To date these collective excitations have been observed by one
other experimental technique, namely
that of tunneling. Using a further set of calculations\cite{Fisher}, it can be
shown that the tunneling density of states is modified, which gives
rise to testable predictions to experimental tunneling I-Vs.  For the
case of the I-V curve of a single nanotube, the model is that there is
a 3d-1d tunneling interface of sorts as the ``ohmic contact'' of one
end of the tube, and a 1d-3d tunneling interface at the other ``ohmic
contact.'' Experiments have observed\cite{BockrathNature} power-law behavior that is
consistent with the tunneling predictions, namely $dI/dV\propto V^{\alpha}$,
where $\alpha = (g^{-1}-1)/4$ or $\alpha=(g^{-1}+g-2)/8$, depending on whether
the contact is at the end or in the bulk of the tube. 
In the 3d-1d tunneling case, an electron
tunnels into the 1d system, which simultaneously excites an infinite
number of 1d plasmons. In reference~\cite{BockrathNature}, experimentally
observed values of $\alpha$ vary between 0.33 and 0.38 for
end-tunneling, and 0.5 and 0.7 for bulk tunneling, giving values of g
between 0.26 and 0.33. A recent paper\cite{Auslaender} also
measured tunneling from one 1d quantum wire in GaAs to another 1d
quantum wire in GaAs. There, they found $g\approx 0.75$. Both of these
approaches are interesting and significant.

In this manuscript we would like to present a different and
complementary method to measure these collective excitations {\it directly},
by exciting them with a microwave (GHz) voltage. In particular, we would
like to measure the wave velocity under a variety of conditions,
including different distances from the nanotube to the ground plane,
to see how the electromagnetic environment effects the properties of
collective excitations in one-dimensional quantum systems.  
An additional capability of the
technique described below would be to measure the 1d plasmon damping, including
dependence on temperature and disorder. This
high-frequency measurement may also have direct applications in
determining the switching speed of a variety of nanotube based 
electronic devices.

\subsection{Characteristic impedance}

Another property of interest of the transmission line is the
characteristic impedance, defined as the ratio of the ac voltage to
the ac current. This is especially important for measurement purposes.
In the circuit model presented above, for a
right-going plasmon wave, the ratio of the ac voltage to the ac
current is independent of position, and is given by:
\begin{equation}
Z_{c,general}=\sqrt{{\mathcal L}\over{\mathcal C}}.
\end{equation}
As we did for the wave velocity, we have to consider the magnetic and kinetic
inductance, as well as the electrostatic and quantum capacitance. Upon
considering the magnetic and electrostatic inductance only, one
recovers the characteristic impedance of free space:
\begin{equation}
Z_{c,free space}=\sqrt{{\mathcal L}_M\over{\mathcal
    C}_{ES}}=\sqrt{\mu\over\epsilon}\equiv Z_0=377~\Omega.
\end{equation}
On the other hand, if one considers only the quantum capacitance and
only the kinetic inductance, the characteristic impedance turns out to
be the resistance quantum:
\begin{equation}
Z_{c,non-interacting,spinless}=\sqrt{{\mathcal L}_K\over{\mathcal C}_Q}={h\over 2e^2}=12.5~k\Omega.
\end{equation}
Now, if one considers the kinetic inductance and both components of
the capacitance (electrostatic + quantum), then one finds:
\begin{eqnarray}
Z_{c,non-interacting,spinless}=\sqrt{{\mathcal L}_K\over{\mathcal C}_{total}}
=\sqrt{{{\mathcal L}_K\over{\mathcal C}_{ES}}+{{\mathcal
      L}_K\over{\mathcal C}_Q}}\nonumber\\
=\sqrt{{\mathcal L}_K\over{\mathcal C}_Q}~\biggl(1+\alpha~{\pi\over
  2}~{c\over v_F}~{1\over ln(h/d)}\biggr)^{-1/2}
=g_{spinless}~{h\over 2e^2},
\end{eqnarray}
where we have inserted the definition of $g_{spinless}$.
This immediately suggests a second method of measuring g at GHz
frequencies, by measuring the characteristic impedance of the
transmission line. We discuss the geometries of interest in detail in
a later section. For now we would like to comment that, even though
the characteristic impedance measurement at high frequencies of high
resistances is challenging, the predicted variation of the
characteristic impedance from the non-interacting $h/2e^2$ is large,
of order 100\%.

To be complete, we must include the magnetic inductance as well,
yielding the full solution to the characteristic impedance:
\begin{eqnarray}
Z_{c,total,spinless}=\sqrt{{\mathcal L}_{total}\over{\mathcal C}_{total}}
= \sqrt{{\bigl({\mathcal L}_K + {\mathcal L}_M\bigr)}\biggl( {1\over
    {\mathcal C}_Q}+{1\over{\mathcal C}_{ES}} \biggr)
}\nonumber \\
={h\over 2e^2}\sqrt{\biggl(1+\alpha{\pi\over 2}{c\over v_F}{1\over
    ln(h/d)}\biggr)
\nonumber
\biggl(1+\alpha{2\over\pi}{v_F\over c}ln(h/d)\biggr)}\\
={h\over 2e^2}\sqrt{1+\alpha\biggl({\pi\over 2}{c\over v_F}{1\over
    ln(h/d)}+{2\over\pi}{v_F\over c}ln(h/d)\biggr)+\alpha^2}
\end{eqnarray}

\subsection{Damping Mechanisms}
\label{sec:damping}
An important question to consider is the damping of the 1d plasma
waves.  Currently very little is known theoretically or experimentally
about the damping {\it mechanisms}. In the absence of such knowledge,
we proceed phenomenologically in the following section.  We model the
damping as distributed resistance along the length of the tube. (This
model of damping of 2d plasmons we recently measured was successful in
describing our experimental results, using the dc resistance to
estimate the ac damping coefficient.) Unfortunately, to date even the
dc resistance of metallic nanotubes is not well quantified. What {\it
  is} known is that the scattering length at low temperatures is at
least 1~$\mu m$, and possibly more. This is known from recent
experiments where the tube length of 1~$\mu m$ gave close to the
Landauer-B\"uttiker theoretical resistance for the dc measurement,
indicating ballistic (scatter free) transport over the length 
of the entire tube\cite{Kong2001}. We state this clearly in an equation for the mean
free path:
\begin{equation}
l_{m.f.p.}>1\mu m
\end{equation}
Now, for dynamical measurements one is usually concerned with the
scattering rate, not length, so if we assume the relationship:
\begin{equation}
l_{m.f.p.}=v_F \tau,
\end{equation}
then we have
\begin{equation}
\label{eq:tau}
\tau > 1 ps
\end{equation}
A separate recent measurement\cite{Hilt} of the mm-wave conductivity
of mats of single walled nanotubes gave a scattering time of 4 ps at room
temperature, but it is unclear how that relates to the scattering
time of individual nanotubes.
The condition that must prevail
for resonant geometries (see below) is that Q must be
greater than one. This implies the condition
\begin{equation}
\omega \tau>1.
\end{equation}
For a 4~ps scattering time, this means the resonant frequency of any
cavity must be greater than 40~GHz. However, we still do not have any
data on how much greater the mean free path is than 1~$\mu m$, and
hence the condition $\omega \tau>1$ could be satisfied at frequencies
below 1 GHz. (In fact an ac measurement of the impedance of a single
nanotube could give more quantitative information about the
mean-free-path as well as the damping coefficient of 1d plasmons.) We speculate that
nanotubes with scattering times satisfying $\omega\tau>1$ at
frequencies below 1~GHz could be grown if they do not already exist; 
this would correspond to a
mean free path of order 100~$\mu m$. We discuss the experimental
consequences of this issue in the next section in more detail.

Another important damping mechanism is if the ground plane is not a
perfect conductor. For a superconducting ground plane, the
approximation of a perfect conductor is a good one. We discuss now two
other cases of interest, that of a metallic film ground plane, and
that of a doped semiconducting ground plane.

A typical deposited metal film will have a thickness of order
0.1~$\mu m$, which is much less than the skin depth at GHz
frequencies. Hence, it can be treated as having a certain sheet
resistance, which is typically of order 1~$\Omega$ per
square at room temperature, although it might be substantially less at
cryogenic temperatures.  For the effective width of order a nanotube width that
participates in the ``grounding'', this would give rise to a
resistance per length ${\mathcal R}$ of the ground plane of order 1~$k\Omega/\mu m$,
which could be a significant source of damping, even if there is no
scattering whatsoever within the nanotube itself. Plasma waves of
frequencies below $1/2\pi\tau={\mathcal R}/2\pi{\mathcal L}=10~GHz$
would be severely damped.
If, instead of a thin film, a bulk metal is used, then the skin depth must be
considered. In that case, the resistance per square must be replaced
by $\rho/\delta_{S.D.}$, where $\rho$ is the bulk resistivity and
$\delta_{S.D.}$ 
the skin depth, which is typically 1~$\mu m$ at 1 GHz in copper
at room temperature. Thus, by increasing
the thickness of the metallic ground plane to 1~$\mu m$, one can
decrease the damping coefficient of the plasmons. However, going any
thicker than the skin depth does not help.  (Interestingly, the exact
same principle applies to gold plating the conductors of coaxial
cables: it is not necessary and certainly not economical to use bulk
gold at rf frequencies for the cable material.) For a 1~$\mu m$ thick
metal ground plane, then, the effective resistance per length that
must be added to the transmission line circuit of the 1d wire can be
of order 100~$\Omega/\mu m$, which is small but not insignificant.

For a doped-semiconductor ground plane, a typical bulk resistivity 
for an n-type doped Si wafer is 10~ohm-cm. 
For this resistivity, the skin depth is of order 1~mm at
1~GHz, so that the effective resistance per square of the ground is
given by 10~$\Omega$-cm/1~mm=100~$\Omega$ per square.  This would give
a resistance per unit length of order 100~k$\Omega/\mu m$, which is a
severe damping, much worse than any scattering in the nanotube
itself. In this case, any plasmons with frequency below
1~THz would be heavily damped.  However, when the skin depth is that large,
corresponding to a distributed resistance in the ``ground'' plane that
continues all the way down to 1~mm below the nanotube, the above
calculations for the characteristic impedance and wave velocity (which
implicitly assumed that the tube length was much larger than the
distance to the ground plane) would have to be revised. We suspect
that further numerical modeling is necessary to fully and
quantitatively understand the
interaction at GHz to THz frequencies between a nanotube and a doped
semiconducting gate, and its effect on damping of 1d Luttinger liquid plasmons.

The important point here is that, even if there is no scattering
whatsoever in the nanotube itself, there may still be damping of the
plasmon mode due to the electromagnetic coupling to the resistive
ground plane.

One final possible loss mechanism is radiation into free space. This was
implicitly neglected in calculating the capacitance using the
electrostatic method\cite{Ramo}. The nano-tube can function as a
nano-antenna, but since the wavelength of the radiation at GHz
frequencies is of order cm, and the tube length is of order $\mu m$, 
it will not be a very efficient nano-antenna, so that radiation losses
are not likely to be significant.

\section{Circuit Model for Metallic Single Wall Carbon Nanotube}
A carbon nanotube, because of its band structure, has two propagating
channels, which we label as channel a and channel b\cite{Dresselhaus}. In
addition, the electrons can be spin up or spin down. Hence, there are
four channels in the Landauer-B\"uttiker formalism language. 
In this section we discuss an effective high-frequency
circuit model which includes the contributions of all four channels,
and makes the spin-charge separation (the hallmark of a Luttinger
liquid) clear and intuitive.

For pedagogical reasons, let us first consider non-interacting spin
1/2 electrons in a
single mode quantum wire at dc.  The current can be carried by either spin
up or spin down electrons.  Usually, when we measure the conductance
of such a wire, the electrical contacts on both ends of the wire are
to both the spin up and spin down channel simultaneously, so that the
effective circuit model is two quantum channels in parallel. However,
if we could inject current in one direction in the spin up channel,
and extract current in the spin down channel, then the net electrical
current (the {\it charge} current) 
would be zero. However, there would be a {\it
  spin} current. This clearly illustrates the separation of
spin-charge currents in a 1d wire at d.c. Below we consider the
generalization to the ac case, and we consider a case where there are
two modes for each spin orientation, correct for a carbon nanotube. We
will neglect the magnetic inductance in what follows. Our approaches
parallels that of reference~\cite{Sonin}, which in turn parallels that
of reference~\cite{Matveev}. We go further than these references,
though, in diagonalizing and calculating the impedance
matrix, and relating this to the {\it measurable} effective circuit impedance of a 1d
plasmon.

\subsection{Non-interacting circuit model for metallic single wall
  carbon nanotube}

\begin{figure}
\epsfig{file=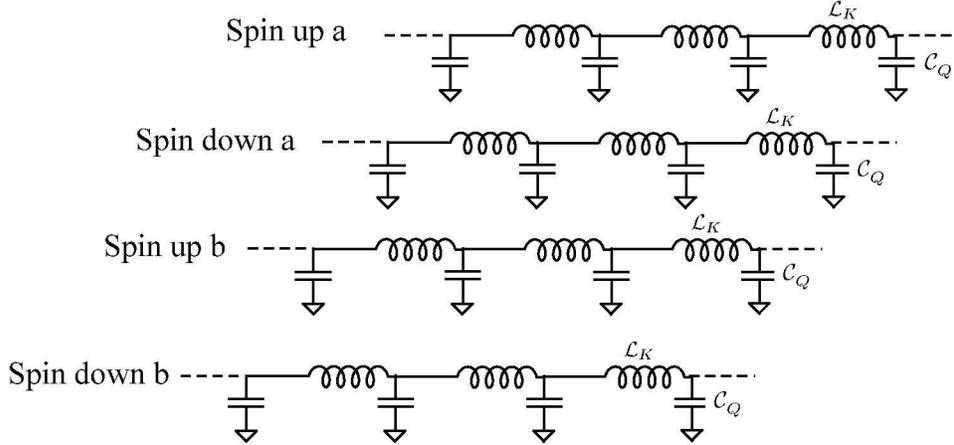}
\caption{Circuit model for non-interacting electrons in a single
  walled carbon nanotube. Each channel (transmission line) is independent of the others.}
\label{fig:paralleltlines}
\end{figure}
The non-interacting ac circuit model of a single-walled carbon nanotube is fairly
straightforward: We simply have four quantum channels in parallel
each with its own kinetic inductance and quantum capacitance per unit
length. (Neglecting the electrostatic capacitance is equivalent to
neglecting the electron-electron interactions.)
All of the above calculations would apply to that system,
accept that there are four transmission lines in parallel. The ends of
all four transmission lines are usually contacted simultaneously by
electrical contacts to SWNTs. (Injecting spin-polarized current into only
the spin up channels has not yet been accomplished experimentally, to
our knowledge.) We draw in figure~\ref{fig:paralleltlines} the effective circuit diagram in
this case.

\subsection{Interacting circuit model for metallic single wall
  carbon nanotube}
\begin{figure}
\epsfig{file=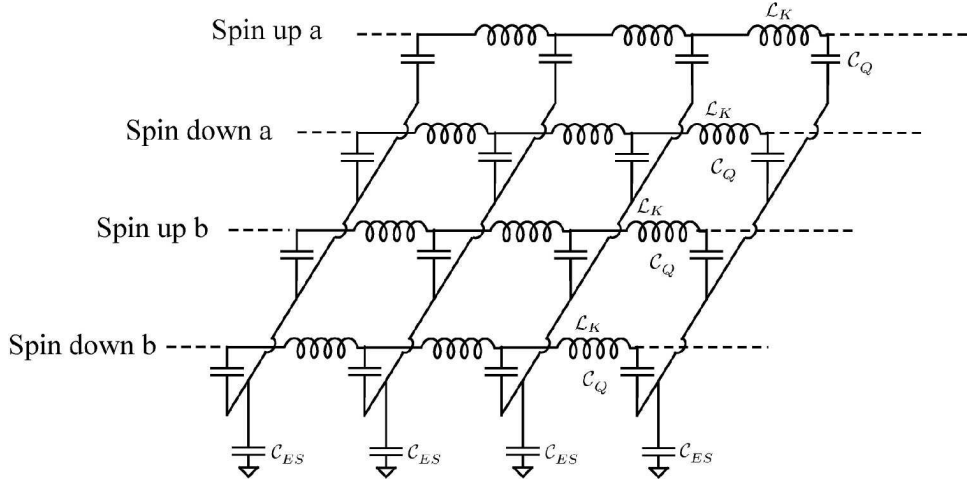}
\caption{AC circuit model for {\it interacting} electrons in a carbon nanotube.}
\label{fig:interactingtlines}
\end{figure}

At this point, we have to take into account the electron-electron
interaction. Apparently this can be done in a phenomenological way by
using the electrostatic capacitance.  The coulomb energy per unit
length is given by
\begin{equation}
E_c={(\rho_{total})^2\over 2{\mathcal C}_{ES}}
={1\over 2{\mathcal C}_{ES}}\biggl(\sum_{i=1}^4 {\rho_i}\biggr)^2
={1\over 2{\mathcal C}_{ES}}\bigl(\rho_{a\uparrow}+\rho_{a\downarrow}+\rho_{b\uparrow}+\rho_{b\downarrow}\bigr)^2
\end{equation}
where $\rho_i$ is the charge per unit length in the ith mode. The
circuit diagram of figure~\ref{fig:interactingtlines} takes this charging energy into account
correctly, and is the central result of this paper.

At this point we have a coupling between the four modes, which is
immediately obvious in the circuit diagram in figure~\ref{fig:interactingtlines}.  Before we
consider the formal mathematics, let us think about physically
meaningful measurements. As in the dc case, if we apply an ac voltage
to the nanotube, we are exciting all four channels
simultaneously. (This is assuming the incoming current is not spin
polarized, another exciting possibility we will not consider in this
manuscript\cite{Balents}.) Therefore, at one end of the nanotube (the
ground end) all four
channels have zero voltage. At the other end of the nanotube (the
``hot'' end), all four channels have the same voltage, $V_0$, say. By
inspection of the circuit diagram, the voltage along the nanotube will
be the same for all four channels. This is actually a normal mode of
the coupled system, namely exciting all channels equally. It should
also be obvious from inspecting the circuit diagram that there is no
spin current in this case: as many spin up electrons move from right to left as
spin down. As we will show below, there are three other normal modes
which do not carry net current. Since they do not
have net current flowing and they are called neutral modes. They do
carry spin currents, though. Hence, the separation between spin and
charge currents, which is one of the hallmarks of a Luttinger liquid.

We now proceed mathematically to solve for the normal modes.
The charge per unit length of the ith mode is related to the voltages of the four
other modes, which (upon inspection of the circuit diagram in
figure~\ref{fig:interactingtlines}) 
we write as a matrix generalization of Q=CV:
\begin{equation}
\left(\matrix{V_{a\uparrow}(x,t)\cr V_{a\downarrow}(x,t)\cr V_{b\uparrow}(x,t)\cr V_{b\downarrow}(x,t)}\right)
=\left(\matrix{
{\mathcal C}_Q^{-1}+{\mathcal C}_{ES}^{-1}&{\mathcal C}_{ES}^{-1}&{\mathcal
  C}_{ES}^{-1}&{\mathcal C}_{ES}^{-1}\cr
{\mathcal C}_{ES}^{-1}&{\mathcal C}_Q^{-1}+{\mathcal C}_{ES}^{-1}&{\mathcal
  C}_{ES}^{-1}&{\mathcal C}_{ES}^{-1}\cr
{\mathcal C}_{ES}^{-1}&{\mathcal C}_{ES}^{-1}&{\mathcal C}_Q^{-1}+{\mathcal
  C}_{ES}^{-1}&{\mathcal C}_{ES}^{-1}\cr
{\mathcal C}_{ES}^{-1}&{\mathcal C}_{ES}^{-1}&{\mathcal
  C}_{ES}^{-1}&{\mathcal C}_Q^{-1}+{\mathcal C}_{ES}^{-1}\cr
}\right)
\left(\matrix{\rho_{a\uparrow}(x,t)\cr \rho_{a\downarrow}(x,t)\cr \rho_{b\uparrow}(x,t)\cr \rho_{b\downarrow}(x,t)}\right)
\end{equation}
(This is equivalent to equation 24 in reference~\cite{Sonin}.) We write this in vector notation as:
\begin{equation}
\label{eq:capacitance}
\vec{V}(x,t)=
{\bf C^{-1}}\vec{\rho}(x,t)
\end{equation}
At this point, we can follow the derivation of the telegrapher
equations, using the matrix generalization. Kirchoff's voltage law gives
\begin{equation}
{\partial \vec{V}(x,t)\over\partial x}=-{\mathcal L}_K {\partial \vec{I}(x,t)\over \partial t}.
\end{equation}
(This can be seen by considering the voltage just to the left and just
to the right of any of the inductors drawn in figure~\ref{fig:interactingtlines}). In the
derivation of the telegrapher equations, Kirchoff's current law is
usually used. It is easier in this case to use the continuity equation, which in 1d is given by:
\begin{equation}
{\partial \vec{\rho}(x,t)\over\partial t}=-{\partial \vec{I}(x,t)\over\partial x}.
\end{equation}
We now proceed to take the second time derivative of equation~\ref{eq:capacitance}, yielding:
\begin{eqnarray}
{\partial\over \partial t}{\partial\over \partial t}\vec{V}(x,t)=
{\partial\over \partial t}{\bf C^{-1}}{\partial
  \vec{\rho}(x,t)\over\partial t}=
-{\bf C^{-1}}{\partial\over \partial t}{\partial \vec{I}(x,t)\over \partial
  x}=\nonumber\\
-{\bf C^{-1}}{\partial\over \partial x}{\partial \vec{I}(x,t)\over \partial
  t}=
{\bf C^{-1}}{\partial\over \partial x}{1\over {\mathcal
    L}_K}{\partial \vec{V}(x,t)\over \partial x}.
\end{eqnarray}
In sum,
\begin{equation}
\label{eq:voltagewave}
{\partial^2\vec{V}(x,t)\over \partial t^2}={{\bf C^{-1}}\over {\mathcal L}_K}{\partial^2\vec{V}(x,t)\over \partial x^2}.
\end{equation}
Using the same methods it can be shown that:
\begin{equation}
\label{eq:currentwave}
{\partial^2\vec{I}(x,t)\over \partial t^2}={{\bf C^{-1}}\over {\mathcal L}_K}{\partial^2\vec{I}(x,t)\over \partial x^2}.
\end{equation}
Thus, we have a set of four coupled wave equations for the voltage and
current on each line. 

Finally, there exists a matrix relating $\vec{V}(x,t)$ and
$\vec{I}(x,t)$, the impedance matrix. This is discussed in reference~\cite{Sonin} in
this basis. We do not consider the impedance matrix in this basis here, as it is not
relevant to the experimental setup we discuss below. In contrast to
reference~\cite{Sonin}, we will discuss the impedance in a different basis,
where it is diagonal.

Let us consider the voltage wave equation, equation~\ref{eq:voltagewave}.  
If ${\mathcal C}_{ES}=0$, then {\bf C} is diagonal, and the voltage
wave (plasmon) in each mode is independent of the others, all moving
at the Fermi velocity. If ${\mathcal C}_{ES}$ is non-zero, this is
tantamount to saying there are interactions, and the four modes are
coupled.  We need now to diagonalize the equations of motion to find
the normal modes. If we want to consider solutions of the form:
\begin{equation}
\label{eq:planewave}
\vec{V}(x,t)=\vec{V}_0e^{i(kx-\omega t)},
\end{equation}
then we must find which values of $\vec{V}_0$ will work solve the
coupled wave equations, eqns.~\ref{eq:voltagewave} and \ref{eq:currentwave}. In other
words, we need to find a set of vectors which diagonalizes the
capacitance matrix.  Specifically, we must solve (on plugging the
above equation~\ref{eq:planewave} into the voltage wave 
equation~\ref{eq:voltagewave}):
\begin{equation}
{\omega^2\over k^2}\vec{V}_0={{\bf C^{-1}}\over{\mathcal L}_K}\vec{V}_0.
\end{equation}
(This is equivalent to equation 27 of reference~\cite{Sonin}, which in
turn is equivalent to equation 11 of reference~\cite{Matveev}.)
The eigenvectors are:
\begin{equation}
\vec{V}_0=\bordermatrix{&C.M\cr
&\matrix{1\cr1\cr1\cr1}\cr},
\bordermatrix{&D1\cr
&\matrix{1\cr1\cr-1\cr-1}\cr},
\bordermatrix{&D2\cr
&\matrix{1\cr-1\cr1\cr-1\cr}\cr},
\bordermatrix{&D3\cr
&\matrix{1\cr-1\cr-1\cr1}\cr}
\end{equation}
We have labeled the eigenvectors C.M. for ``common mode'' and D1-D3
for differential 1-3. The ``common mode'' vector is the fundamental
charged excitation in a Luttinger liquid. Below we discuss a method to
excite these modes with a microwave voltage. The other three are neutral,
that is they carry no net (charge) current. (Since the other three are
degenerate, it is possible to chose a different basis for the other
three. A basis of non-orthogonal degenerate eigenvectors was used in reference~\cite{Sonin},
but we chose the orthogonal eigenvectors as in
refs.~\cite{Egger1,Egger2,Kane}.) 
However, the differential
modes do carry {\it spin} current. These are the neutral and
charged modes of a Luttinger liquid.  
This is the clear separation of spin and charge degrees of
freedom which is the hallmark of a Luttinger liquid.

In the new basis, the capacitance matrix is diagonal. If we write the
voltages in the new basis as
\begin{equation}
\vec{V}'_0(x,t)=\left(\matrix{V_{C.M.}(x,t)\cr V_{D1}(x,t) \cr V_{D2}(x,t) \cr V_{D3}(x,t)}
\right)=
\left(\matrix{
V_{a\uparrow}(x,t)+V_{a\downarrow}(x,t)+V_{b\uparrow}(x,t)+V_{b\downarrow}(x,t)\cr
V_{a\uparrow}(x,t)+V_{a\downarrow}(x,t)-V_{b\uparrow}(x,t)-V_{b\downarrow}(x,t)\cr
V_{a\uparrow}(x,t)-V_{a\downarrow}(x,t)+V_{b\uparrow}(x,t)-V_{b\downarrow}(x,t)\cr
V_{a\uparrow}(x,t)-V_{a\downarrow}(x,t)-V_{b\uparrow}(x,t)+V_{b\downarrow}(x,t)\cr
}\right),
\label{eq:changeofvariables}
\end{equation}
and similarly for $\vec{\rho}'_0(x,t)$ and $\vec{I}'_0(x,t)$, then the new
capacitance matrix is simply given by:
\begin{equation}
\left(\matrix{V_{C.M.}(x,t)\cr V_{D1}(x,t)\cr V_{D2}(x,t)\cr V_{D3}(x,t)}\right)
=\left(\matrix{
{\mathcal C}_Q^{-1}+4{\mathcal C}_{ES}^{-1}&0&0&0\cr
0&{\mathcal C}_Q^{-1}&0&0\cr
0&0&{\mathcal C}_{Q}^{-1}&0\cr
0&0&0&{\mathcal C}_{Q}^{-1}}\right)
\left(\matrix{\rho_{C.M.}(x,t)\cr \rho_{D1}(x,t)\cr \rho_{D2}(x,t)\cr \rho_{D3}(x,t)}\right)
\end{equation}
or, in vector notation,
\begin{equation}
\vec{V}'(x,t)={\bf C'^{-1}}\vec{\rho}'(x,t)
\end{equation}
Additionally, in the new basis the following holds:
\begin{equation}
\label{eq:dvdz}
{\partial \vec{V}'(x,t)\over\partial x}=-{\mathcal L}_K {\partial \vec{I}(x,t)'\over \partial t}.
\end{equation}
and:
\begin{equation}
{\partial \vec{\rho}'(x,t)\over\partial t}=-{\partial \vec{I}'(x,t)\over\partial x}.
\end{equation}
In this new basis, the wave equation for the voltage is now diagonal,
with new wave equations given by:
\begin{eqnarray}
{\partial^2 V_{CM}(x,t)\over \partial t^2}={1\over{\mathcal L}_K}
\biggl({1\over{\mathcal C}_Q}+{4\over{\mathcal
    C}_{ES}}\biggr){\partial^2 V_{CM}(x,t)\over \partial x^2}\\
{\partial^2 V_{D1}(x,t)\over \partial t^2}
={1\over{\mathcal C}_Q{\mathcal L}_K}{\partial^2 V_{D1}(x,t)\over \partial x^2},
\label{eq:diffvoltagewave}
\end{eqnarray}
with the equation for D2 and D3 the same as for D1. In vector form:
\begin{equation}
{\partial^2\vec{V}'(x,t)\over \partial t^2}={{\bf C'^{-1}}\over {\mathcal L}_K}{\partial^2\vec{V}'(x,t)\over \partial x^2}.
\end{equation}
Similarly, one can show that:
\begin{equation}
{\partial^2\vec{I}'(x,t)\over \partial t^2}={{\bf C'^{-1}}\over {\mathcal L}_K}{\partial^2\vec{I}'(x,t)\over \partial x^2}.
\end{equation}
Now, the wave velocity for the differential modes is just the Fermi velocity (using
equation~\ref{eq:diffvoltagewave} above).  However, the velocity for the common mode,
i.e. 1d plasmon, is given by:
\begin{equation}
\label{eq:vplasmon}
v_{p} = \sqrt{{1\over {\mathcal L}_K}\biggl({1\over{\mathcal C}_Q}+
    {4\over {\mathcal C}_{ES}}\biggr)}
=v_F \sqrt{1+{4
    {\mathcal C}_Q\over {\mathcal C}_{ES}}}\equiv v_F/g
\end{equation}
This equation (which is not a new result\cite{Fisher}) defines g for a SWNT.
Now, let us consider solutions to the voltage and current wave
equations in the diagonal basis, in order to determine the
characteristic impedance. Since the wave equations are diagonal
(i.e. uncoupled), if we can excite the common mode, none of the other
modes will be excited.

The general solutions are of the form:
\begin{equation}
\label{eq:vcm}
V_{C.M.}(x,t)=V_{C.M}^{+}e^{i(kx-\omega t)} + V_{C.M.}^{-} e^{i(-kx+\omega t)},
\end{equation}
and
\begin{equation}
\label{eq:current}
I_{C.M.}(x,t)=I_{C.M}^{+}e^{i(kx-\omega t)} + I_{C.M.}^{-} e^{i(-kx-\omega t)}.
\end{equation}
Applying equation~\ref{eq:dvdz} to the above equation~\ref{eq:current} for the
current gives the following:
\begin{equation}
I_{C.M.}(x,t)=\sqrt{{4{\mathcal L}_K\over{\mathcal C}_{ES}}+{{\mathcal
      L}_K\over{\mathcal C}_Q}}
\biggl(V_{C.M}^{+}e^{i(kx-\omega t)} - V_{C.M.}^{-} e^{i(-kx+\omega t)}\biggr)
\label{eq:current2}
\end{equation}
The ratio of the ac voltage to the ac current on the line is defined
as the ``characteristic impedance'', which can be seen from comparing
equation~\ref{eq:current2} to equation~\ref{eq:vcm}. Thus, for the common mode, the
characteristic impedance is given by:
\begin{equation}
\label{eq:zchar}
Z_{c,C.M.}\equiv{V_{C.M.}^{+}\over I_{C.M.}^{+}}=-{V_{C.M.}^{-}\over I_{C.M.}^{-}}=
\sqrt{{4{\mathcal L}_K\over{\mathcal C}_{ES}}+{{\mathcal
      L}_K\over{\mathcal C}_Q}}
={1\over g}~{h\over 2 e^2}
\end{equation}
This is a very important number which will be used in the experimental
techniques section to be discussed below.
Our result differs from equation 37 of reference~\cite{Sonin}
because we are considering the excitation of only the common mode,
i.e. Luttinger liquid charge mode. Reference~\cite{Sonin} considered the
excitation of mode $V_{a\uparrow}$, i.e. a superposition of charge and spin modes. Below
we discuss how our method excites only the charge mode, and not the
spin mode, so that our calculation is more germane to our experimental
technique described below to directly excite Luttinger liquid
collective modes.

Now, it is important to realize that what one measures is not exactly
$Z_c$ for the common mode. The common mode impedance is the sum of the
voltages ($V_{a,\uparrow}+V_{a,\downarrow}+V_{b,\uparrow}+V_{b,\downarrow}$)
divided by the sum of the 
currents
($I_{a,\uparrow}+I_{a,\downarrow}+I_{b,\uparrow}+I_{b,\downarrow}$). 
The sum of the currents
is what flows into an external circuit.  However, when coupled to an
external circuit all of the voltages are equal to the eternally
measured voltage, so that the common mode voltage is actually 4 times
larger than the voltage measured at the end of the tube by an external circuit.
That is why our equation~\ref{eq:zchar} differs from equation~3 of 
reference~\cite{Tarkiainen}.

Finally, for the sake of completeness, it can be shown that the 
following is the characteristic impedance of the other three modes modes:
\begin{equation}
Z_{c,D1}\equiv{V_{D1}\over I_{D1}}=\sqrt{{\mathcal L}_K\over{\mathcal
    C}_{Q}}={h\over 2 e^2}
\end{equation}
This describes the ratio of the voltage to the current when the
spin-wave is excited.

\section{Measurement technique}
In this section we consider various methods of exciting the common
mode (charged) Luttinger liquid plasmon with an ac voltage.  In order
to describe this, let us first consider measurements of the dc
conductance of a single walled carbon nanotube. In the experiments
performed to date, current flows through all four channels. In the
case of tubes which approach $e^2/4h$ of conductance, i.e. where the
macroscopic ``lead'' contacts all four channels, the current is
equally distributed among all four channels. This is equivalent to
exciting only the common mode current, and the common mode voltage as
well. We would like to describe below a set of experiments where we
contact all four channels simultaneously with an {\it ac}
(microwave) voltage. This finite-frequency measurement will excite
only the common mode (charged) Luttinger liquid 1d plasmon.  (We do
not discuss it here, but it would also be interesting to drive a
microwave spin polarized current to excite the spin modes of the
Luttinger liquid. Currently we are unaware of any experimental
technique to do that.)  Since there is a finite frequency, there will
also be a wave-vector introduced. If we measure the frequency
dependent impedance of the nanotube, we should be able to determine
the frequency at which there are one, two, three... standing waves in
the tube, and hence measure the dispersion curve and wave velocity of
the 1d plasmon. From equation~\ref{eq:vplasmon}, this will allow a direct
measurement of the parameter g.

\begin{figure}
\epsfig{file=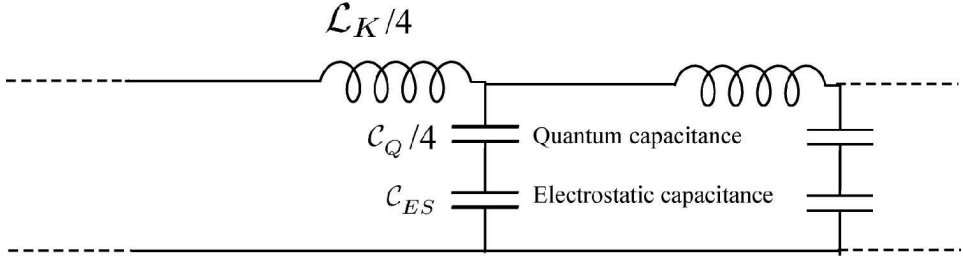}
\caption{Charge mode (``common mode'') effective circuit diagram.}
\label{fig:commonmodecircuit}
\end{figure}

Above, we derived a set of differential equations describing the
current and voltage for all four modes in a Luttinger liquid 
(3 neutral spin waves and one charge wave). 
Now, we would like to consider only the charged
mode, and calculate the effective, frequency dependent impedance that
one would expect for a carbon nanotube at microwave frequencies.

At this point, we have two options: First, we can continue to work with the
circuit diagram in figure~\ref{fig:interactingtlines}, and apply the
appropriate boundary conditions for the measurement geometries that
we will consider below. This has the advantage that all four
channels are still present in our effective circuit model, but it is
somewhat complicated. However, for the boundary conditions this is
actually a simpler option, as we will see. 

Option two is to use the fact that we are
considering only exciting the common mode in this paper, and to
replace figure~\ref{fig:interactingtlines} with an ``effective'' circuit
diagram consisting of a {\it single} transmission line with rescaled inductance
and capacitance per unit length. This is indicated in
figure~\ref{fig:commonmodecircuit},
where the effective inductance per unit length is
now ${\mathcal L}_K/4$, and the effective capacitance per unit length
is given by $({\mathcal C}_{ES}^{-1}+{4\mathcal C}_Q^{-1})^{-1}$. 
The wave velocity of this ``effective'' circuit model is the same as the
wave velocity of the common mode (given by equation~\ref{eq:vplasmon}). The
characteristic impedance of this effective circuit model is 1/4 of the
characteristic impedance of the common mode (given by
equation~\ref{eq:zchar}), which is due to the following:  When we excite
the common mode voltage, all four voltages
($V_{a,\uparrow},V_{a,\downarrow},V_{b,\uparrow},V_{b,\downarrow}$) are
{\it equal}, so that the common mode voltage $V_{C.M.}$ is four times
larger than the measured voltage by the external circuit, since
$V_{C.M.}=V_{a,\uparrow}+V_{a,\downarrow}+V_{b,\uparrow}+V_{b,\downarrow}$
as given in equation~\ref{eq:changeofvariables}. (The common mode current is that same
as the measured current.) The advantage of using the circuit diagram
proposed in figure~\ref{fig:commonmodecircuit} is that we only have to
deal with {\it one} transmission line. The disadvantage is that the
effective boundary conditions for the geometries we consider below are
not obvious and require careful consideration. In the following
sections we will use both descriptions, according to convenience and
relevance to the particular boundary conditions under consideration.

We proceed in this section as follows:
We first consider an ``ohmically'' contacted nanotube, by which we
mean tubes with dc electrical contacts with perfect transparency which
have $4e^2/h$ of conductance. Of course, this is a linearized model of
the dc resistance, which can have a significant non-linear
current-voltage relationship. It is beyond the scope of this
manuscript to include non-linear resistances in the effective 
circuit impedance. After considering ``ohmically'' contacted
nanotubes, which are not trivial to achieve technologically, 
we consider a capacitively
contacted nanotube which does not require dc contact. Such a
measurement geometry should be much easier to achieve, since in
essence it only requires evaporating a metal lead onto a nanotube.
A discussion of the measurement geometries requires 
careful consideration of the boundary
conditions for the 1d plasmons, which we treat below.

\subsection{Ohmic contacted measurement}

\begin{figure}
\epsfig{file=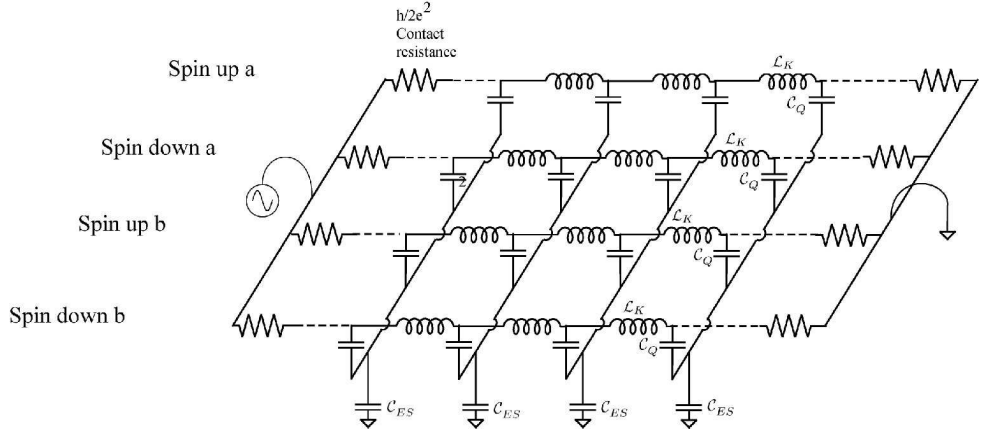}
\caption{Circuit diagram for a SWNT with dc electrical
  contacts at both ends.}
\label{fig:zintcontact}
\end{figure}

We begin by considering the simplest measurement geometry, that of an
``ohmically'' contacted single wall nanotube with perfect transparency
at both ends.  The d.c. conductance is just $4 e^2/h$, since there are
two channels and two spin orientations per channel.  Tubes with dc
resistance approaching this value have recently been
fabricated\cite{Kong2001}. 
For ac (dynamical) impedance measurements, 
we really do not know where to put the contact resistance in the ac
circuit diagram. Experimentally, the high frequency conductivity 
of nano-scale systems is an unexplored regime
of mesoscopic physics; there have been few
experiments\cite{BurkeAPL,BurkePRB,Alex,Schoelkopf1997,Pieper1994}. 
We speculate that the impedance can be modeled as a ``contact''
resistance, which is discussed more rigorously in
reference~\cite{Blanter,Buttiker:1997}. Following
reference~\cite{Blanter},
we model the contact resistance as split into a
contact resistance ("charge relaxation resistance"\cite{Buttiker1993}) 
on {\it each} side of the wire, so that each
side has 1/2 of the total dc resistance. To be explicitly clear, we
draw in figure~\ref{fig:zintcontact} the circuit diagram we are proposing.
At dc, each of the four channels has $h/2e^2+h/2e^2=h/e^2$ of total contact resistance.
Since there are four quantum channels in parallel, the total
resistance is given by $h/4e^2$, the Landauer-B\"uttiker expected value.

\begin{figure}
\epsfig{file=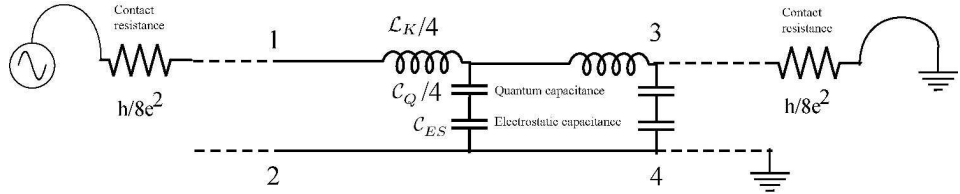}
\caption{{\it Effective} circuit diagram for a SWNT
with dc electrical contacts at both ends.}
\label{fig:tube2}
\end{figure}

Now, it is possible to define an effective circuit diagram along the
lines of figure~\ref{fig:interactingtlines}. We show in figure~\ref{fig:tube2} the
effective circuit which is the ``Norton equivalent'' circuit to
figure~\ref{fig:zintcontact}.  The values for the contact resistance
are shown as $e^2/8h$ each. It is obvious from the circuit diagram
that the dc resistance is equal to $e^2/4h$, so that our model is
correct in the dc limit.

Before we continue there is one more issue that needs to be
discussed.  That is the issue of damping along the length of the
tube.  We again speculate that the dc resistance per unit length gives
information about the distributed damping of the 1d plasmons.  We
model this as a distributed resistance per unit length ${\mathcal
  R}$. We must again be careful about the factor of four when we
define this parameter. In our nomenclature, we define ${\mathcal R}$
as the dc resistance per unit length of all four channels in parallel.
 Of course, according to the scaling theory of
localization\cite{sheng}, the resistance of a 1d system is expected to
scale exponentially with length on the length scale of the
localization length. However, it is known experimentally that the
localization length is greater than a few $\mu m$, but it is not known
how long the localization length really is. Our simplified model of a resistance
per unit length violates the expected (but never observed) exponential
scaling of the resistance with length in 1d, but makes the problem tractable.
According to our definition of ${\mathcal R}$ as the resistance per
length of all four channels in parallel, we must insert a distributed
resistance of $4{\mathcal R}$ into each of channels in 
figure~\ref{fig:zintcontact}. Or, equivalently, we
must insert a resistance per unit length of ${\mathcal R}$ in the
effective circuit diagram figure~\ref{fig:tube2}. Our discussion of
damping in section~\ref{sec:damping} is consistent with this definition.
We will consider various numerical values of ${\mathcal R}$ below.

At this point we are in a position to calculate the (complex, frequency
dependent) ratio of the ac voltage
to the ac current entering the left end of the nanotube, the {\it
  impedance}. We do this by ``mapping'' the problem on to well-known
problems in transmission line theory\cite{Pozar}. We proceed in
two steps: First, we consider the impedance
without the contact resistance on the left hand side. In other words,
we calculate the impedance from point 1 in figure~\ref{fig:tube2} to
ground.
This is equal to the impedance from point 1 to point 2 in
figure~\ref{fig:tube2}, which is 
equivalent to the input impedance of a (possibly lossy)
transmission line with
characteristic impedance $Z_{c,effective}$, which is ``terminated'' by a
``load''impedance $Z_L$ which in this case is simply the contact
resistance, i.e. $Z_L = R_{contact}=h/8e^2$. This is a standard result
in microwave theory, which we repeat here for convenience:
\begin{equation}
Z_{in}=Z_{c,effective} {1+\Gamma e^{-2\gamma l}\over 1-\Gamma e^{-2\gamma l}},
\label{eq:zinputgeneral}
\end{equation}
where $l$ is the length of the tube, and $\gamma$ is the propagation constant
of the 1d plasmon, given by:
\begin{equation}
\gamma\equiv \sqrt{({\mathcal R}+i\omega{\mathcal L}_{eff})(i\omega{\mathcal C}_{eff})},
\label{eq:gamma}
\end{equation}
and where we have to defined $Z_{c,effective}$ as
\begin{equation}
Z_{c,effective}\equiv\sqrt{{\mathcal R}+i\omega {\mathcal L}_{eff}\over i\omega {\mathcal C_{eff}}},
\label{eq:zccomplex}
\end{equation}
and where we have defined a new symbol $\Gamma$ (the
reflection coefficient of the plasmon wave off of the right end of
load impedance ``terminating'' the nanotube) as:
\begin{equation}
\Gamma\equiv {Z_L-Z_{c,effective}\over Z_L+Z_{c,effective}}.
\label{eq:biggamma}
\end{equation}
The effective inductance per unit length is:
\begin{equation}
{\mathcal L}_{eff}\equiv {\mathcal L}_K/4,
\end{equation}
and the effective capacitance per unit length is:
\begin{equation}
{\mathcal C}_{eff}^{-1}\equiv4{\mathcal C_Q}^{-1}+{\mathcal C}_{ES}^{-1},
\end{equation}
as we have already indicated in figure~\ref{fig:tube2}.
In the high frequency limit ($\omega>{\mathcal L}_{eff}/{\mathcal
  R}$), $\gamma$ is just the wave-vector k, i.e.
\begin{equation}
\lim_{\omega>{{\mathcal R}/{\mathcal L}_{eff}}}(\gamma)=k\equiv {2\pi\over \lambda}={\omega\over v_{p}}
\end{equation}
\begin{equation}
\lim_{\omega>{{\mathcal R}/{\mathcal
      L}_{eff}}}(Z_{c,effective})={1\over g}{h\over 8e^2}={1\over 4}Z_{c,C.M.}
\label{eq:zclimit}
\end{equation}
For the circuit model where we terminate the end of the transmission
line with a contact impedance equal to half the total dc resistance,
we assume that each of the four transmission lines has a resistance at
each end equal to $h/2e^2$. Therefore, by the definition of the common
mode transmission line parameters, we need to use a load impedance of
$Z_L=h/8e^2$ in equation~\ref{eq:biggamma} in order to implement the
model discussed in the first paragraph of this section.

In the second step of the calculation, we note that the total
impedance is just the contact resistance of the left hand side of the
nanotube plus the input impedance of equation~\ref{eq:zinputgeneral}. If we
take the contact resistance on each side to be half of the total
d.c. resistance (i.e. $R_{contact}=h/8e^2$), then we have the desired result:
\begin{equation}
Z_{nanotube}={h\over 8 e^2}+Z_{c,effective} {1+\Gamma e^{-2\gamma l}\over 1-\Gamma e^{-2\gamma l}}.
\label{eq:znanotube}
\end{equation}
This is a clear prediction that can be experimentally measured.
While it may seem like a complicated result, it is actually quite
elegant. What's more, we recently verified experimentally 
the 2d analog of equation~\ref{eq:znanotube} in reference~\cite{BurkeAPL}.

Before we turn to a numerical evaluation of
equation~\ref{eq:znanotube}, let us consider qualitatively the
expected frequency dependent behavior. At low frequencies, we should
recover the dc limit of a real impedance of $h/4e^2$. This can
indeed be shown to be the case, by taking the $\omega\rightarrow 0$
limit of equation~\ref{eq:znanotube}. As the frequency is
increased (assuming the damping is not too severe, see below), 
there will be resonant peaks in $Z_{nanotube}$ as a function of frequency,
corresponding to first, second, third, etc. harmonic of the
fundamental wavevector set by the finite length of the tube. 
Applying this high frequency voltage would {\it directly} excite the
1d Luttinger liquid low energy excitations (the 1d plasmons). The
locations of these peaks in frequency space can be used to determine
the wave velocity of this mode, and hence g.

At this point the best way to proceed is to evaluate
equation~\ref{eq:znanotube} 
numerically for some possibly typical cases, which leads
into the discussion of the numerical value of the 
distributed resistance ${\mathcal R}$ which (in addition to the
contact resistance) causes damping.
This discussion must be somewhat speculative, 
since the 1d plasmon damping has never been
measured, in fact the 1d plasmon itself has not yet been directly observed.
Currently
very little is known about possible mechanisms. Our model of a
distributed resistance per unit length
gives rise to an exponential decay in a propagating 1d plasmon wave,
with a decay length given by:
\begin{equation}
\label{eq:ldecay}
l_{decay}={2 Z_{c,effective}\over {\mathcal R}}.
\end{equation}
(We implicitly assume the limit $\omega>{\mathcal R}/{\mathcal L}_{eff}$ in
equation~\ref{eq:ldecay}). The more general case will be discussed below.)
Before we discuss estimates for the numerical value of ${\mathcal R}$, 
we discuss what effect it would have on the plasmon resonance discussed above.
As microwave engineers
intuitively know, when the length of the transmission line (in this
case nanotube) is much
longer that the decay length $l_{decay}$, there is no resonant behavior
to the transmission line, and the input impedance becomes the
characteristic impedance of the transmission line $Z_c$, independent of
the ``load'' impedance. Physically, this is because the wave that
propagates toward the load gets essentially completely attenuated
before it reaches the load.  On the other hand, if the transmission
line is shorter than the decay length $l_{decay}$, then the impedance
becomes resonant as in the case we discussed above, with some damping
hence finite Q. 

In the absence of either theory or data, we conjecture that the decay
length scale for 1d Luttinger liquid plasmons
must be at least as long as the mean free path determined from
dc transport measurements. Since the mean free path is known to be at
least 1~$\mu m$ long, the resistance per length is less than
25~k$\Omega/\mu m$ (using equation~\ref{eq:ldecay}).
Another technique to estimate an upper limit on ${\mathcal R}$ is to
use data from recent STM experiments\cite{Fuhrer} 
which measure the voltage drop along the length of the
tube for {\it semiconducting} tubes. There the resistance per unit length is
found to be 9~k$\Omega/\mu m$. (Presumably metallic tubes have an even
lower resistance per unit length.) In this (presumably worst
case) scenario, the damping length $l_{decay}$ would be equal to 
roughly 3~$\mu m$. We consider
below two important cases in turn: first, where the tube length is less than
$l_{decay}$, and second in the ``overdamped'' limit where the tube
length is greater than $l_{decay}$.

For the case of tube lengths less than the decay length, we discuss
nanotubes of length 100~$\mu m$. Recent progress on CVD
growth\cite{Dai} has made such long, SWNTs possible.  
With such a long length, the
resonance frequencies will be in the GHz range, where experiments are
feasible. In the THz frequency range, it
should also be possible to measure frequency-dependent 
properties\cite{PeraltaAPL,PeraltaThesis}, which would be relevant for
tubes with lengths in the $\mu m$ range. 
The technical challenges in the THz range are not straightforward,
though, and generally more difficult than in the GHz range.
Since we have had excellent experimental success with
measuring 2d plasmons\cite{BurkeAPL} in the GHz range, that is where
we focus our attention. However, our predictions should also apply to
THz resonance frequency experiments.

\begin{figure}
\epsfig{file=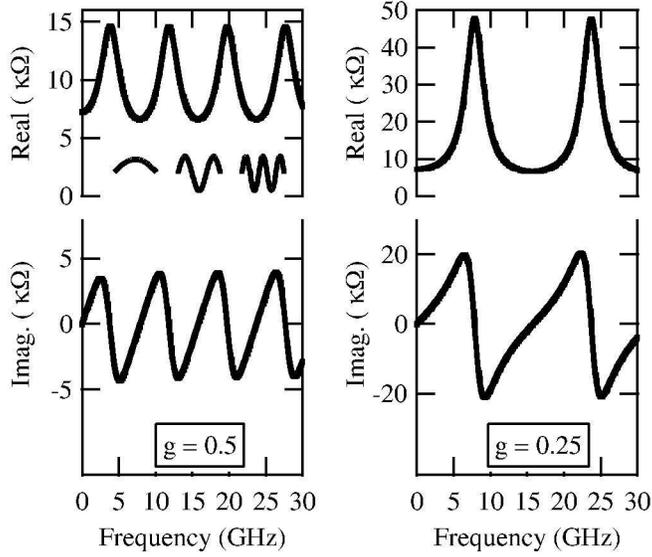}
\caption{Predicted nanotube dynamical impedance for ohmic contact, for two different
  values of g. We assume $l=100~\mu m$.}
\label{fig:ohmic}
\end{figure}

We chose (optimistically) a resistance per unit length of 10~$\Omega/\mu m$,
which is much less than the experimental upper limit of
10~k$\Omega/\mu m$.  In the case that the total resistance distributed
along the length of the nanotube (i.e. ${\mathcal R}*l$) is less than
the contact resistance, the
resistance of the contacts is the dominant damping mechanism. This is
the case for the parameters we have chosen. We plot in
figure~\ref{fig:ohmic} the predicted nanotube dynamical impedance, for
two different values of g. The predicted value of g is 0.25, and we
also plot the predicted value of $Z_{nanotube}$ for g=0.5, which we
achieve by numerically adjusting $C_{ES}$ in our model.

In principle it should be possible to build a measurement apparatus
that could measure this prediction. There are two main technical
challenges: First, the impedance is high, which is difficult for
microwave experiments to resolve. This issue could be solved by
measuring many nanotubes of the same length in parallel, although one
would need to assume that each tube had the same g factor, damping,
etc. The second challenge is that the macroscopic lead will have a
finite capacitance to ground, just by virtue of the fact that the lead
is finite in size.  This capacitance to ground in many
conceivable geometries will provide a low-impedance path to ground in
parallel with the high-impedance nanotube, which will effectively
short the nanotube to ground. This second difficulty makes the
``ohmically'' contacted geometry very difficult to realize experimentally.
However, with sufficient effort it should be feasible.

An interesting prediction of our model is the
frequency at which the first resonance occurs. The real part of the
impedance peaks at a quarter wavelength. 
(It is a general result from microwave
and rf engineering the quarter wavelength structures transform open
circuits to short circuits and vice versa. This fact is used in many
modern rf circuits.) The resonance frequency can be written as:
\begin{equation}
f_{resonance}={v_F\over 4~L}{1\over g}.
\end{equation}
At this frequency, the imaginary part of the impedance crosses zero.
Therefore, if a measurement scheme can be devoiced to measure the where
the imaginary part of the ohmically contacted nanotube impedance
changes sign, this would be a {\it direct} measurement of the Luttinger
liquid parameter g, since L and $v_F$ would be known. An additional
interesting parameter is the equation of the Q of the resonance. This
can be estimated as:
\begin{equation}
\label{eq:Q}
Q={Z_{c,effecitive}\over 2 R_{total}},
\end{equation}
where $R_{total}$ is the total resistance of the nanotube.

\begin{figure}
\epsfig{file=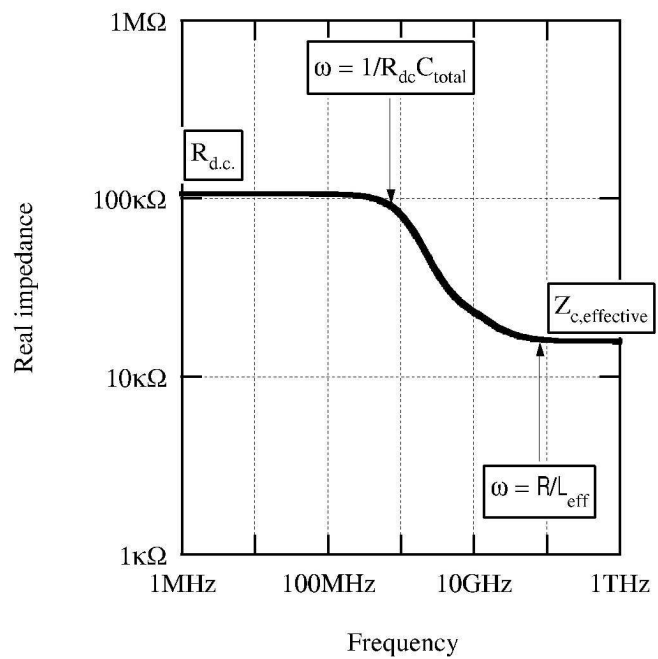}
\caption{Predicted nanotube dynamical impedance in overdamped case.}
\label{fig:znanotubefigureoverdamped}
\end{figure}

We now consider the opposite case, that of an ``overdamped'' 1d
plasmon.  We consider again a tube of length 100~$\mu m$, and now we
consider resistance per unit length of 1~k$\Omega/\mu m$.  In this
case there will be no resonant frequency behavior. We plot in 
figure~\ref{fig:znanotubefigureoverdamped} the
predicted real impedance (using equation~\ref{eq:znanotube}) for these
parameters, assuming g=0.25. There are two qualitative features that
we would like to discuss. First, at dc the real impedance is simply
the resistance per length times the length, i.e. ${\mathcal R}l$.  As
the frequency is increased, the impedance falls. The frequency scale
at which the impedance starts to change is given by the inverse of the
total capacitance (${\mathcal C}_{eff}l$) times the total
resistance. At very high frequencies, the impedance becomes equal to
the effective characteristic impedance given in
equation~\ref{eq:zccomplex}. The frequency at which this occurs is given by
the inverse of the effective ``LR'' time constant, which is the
resistance per unit length divided by the inductance per unit
length. We note that in this high frequency limit, the effective
characteristic impedance ($Z_{c,eff}$) given by equation~\ref{eq:zccomplex}
is mostly real. Therefore, even in the overdamped case where there is
no resonant behavior, the transmission-line behavior of the nanotube
becomes important at frequencies below 1~GHz.

\subsection{Ohmic contacted resonance measurement on one end}

\begin{figure}
\epsfig{file=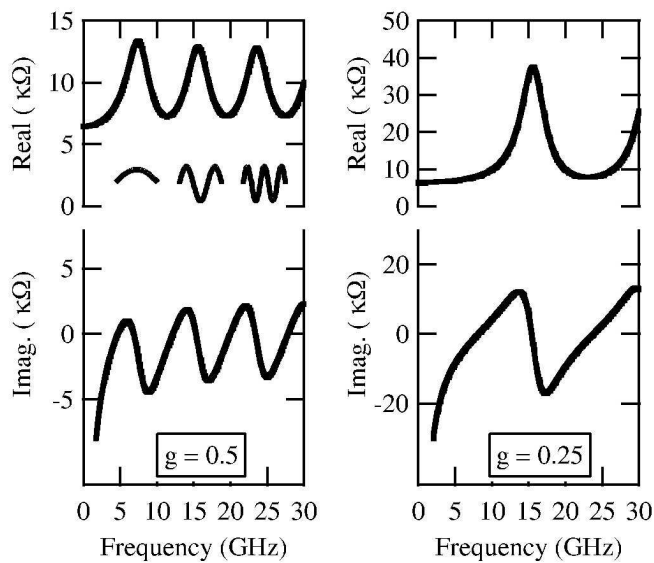}
\caption{Predicted nanotube dynamical impedance for ohmic contact on one end only for two
  different values of g.}
\label{fig:znanotubefigure2}
\end{figure}

Another possible measurement setup would consist of making electrical
contact on one end only of the nanotube, and letting the other end
``float''.  This would correspond to cutting the wire to ground on the
right hand side of figure~\ref{fig:tube2}.  
At dc, no current would flow
so the impedance would be infinite.  However, at ac current could flow
in and out of the end of the tube (charging and discharging the
capacitors), so it is still meaningful to
consider the dynamical impedance.
In this case, we can still use equation~\ref{eq:znanotube} to predict this
dynamical impedance, with a ``load'' impedance in
equation~\ref{eq:biggamma} of infinity (corresponding to an open circuit at
the other end of the nanotube.)  We plot in
figure~\ref{fig:znanotubefigure2} the predicted dynamical impedance in
this case, where we have again assumed a length of 100~$\mu m$, but
where we use a resistance per unit length of 100~$\Omega/\mu m$.
Resonant behavior is still predicted, but now the first peak in the
real impedance occurs at half a wavelength.

\subsection{Capacitively contacted measurement}

The fabrication of electrical contacts to carbon nanotubes with low
resistance at {\it dc} is not a trivial challenge. Even if it can be
achieved, the ``contact'' resistance at ac may be different than it is
at dc for unknown physics reasons. An alternative approach would be to
use capacitive contacts to the nanotube. In the context of the above
discussion, it should be clear that there is already capacitive
coupling between the ground plane and the nanotube, so how can one
achieve capacitive coupling to a macroscopic lead?

\begin{figure}
\epsfig{file=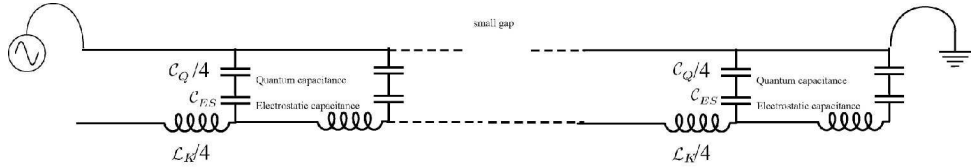}
\caption{Circuit diagram for capacitively coupled nanotube.}
\label{fig:tube3}
\end{figure}

\begin{figure}
\epsfig{file=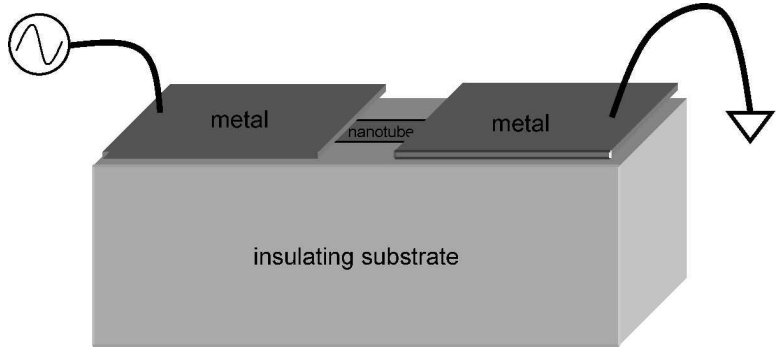}
\caption{Geometry for capacitive contact. The spacing between the
  metal electrodes has been enlarged for clarity. No d.c. electrical
  contact to the nanotube is implied in this picture, only capacitive
  coupling to the leads.}
\label{fig:capgeom}
\end{figure}

One solution is simply to turn the problem upside down.  We envisage
laying a carbon nanotube on an {\it insulating} substrate, and then
evaporating a metallic, macroscopic lead onto the top of one end of
the nanotube, and another macroscopic, metallic lead onto the top of
the other end of the nanotube. One lead is connected to ground, and
the other lead is connected to an ac voltage source.  The impedance
from one lead to the other is measured. This corresponds to measuring
the impedance from one lead to the nanotube plus the impedance from
the nanotube to the other lead. By the symmetry in the problem, we
only need to consider one of those impedances and multiply by two. The
effective circuit diagram we consider is shown in figure~\ref{fig:tube3}.
The physical geometry is indicated schematically in figure~\ref{fig:capgeom}.
This capacitive coupling scheme is exactly the scheme we used for
capacitive coupling to 2d plasmons; 
see our figure~2 in reference\cite{BurkeAPL}.


\begin{figure}
\epsfig{file=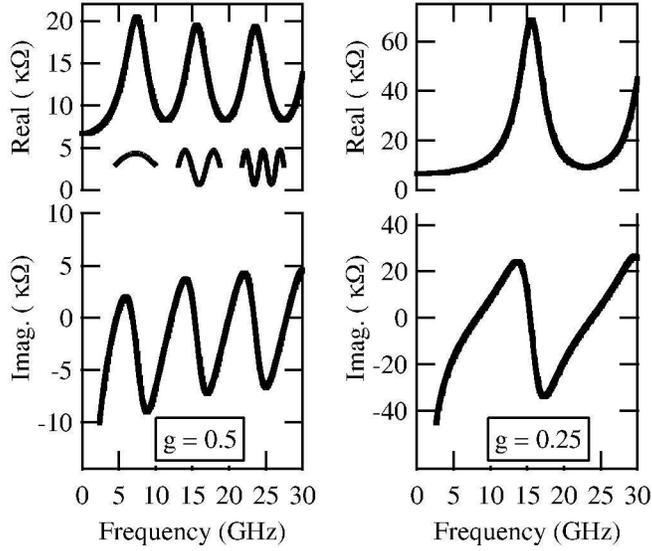}
\caption{Predicted nanotube dynamical impedance for capacitive contact on one end for
  two different values of g.}
\label{fig:znanotubefigure6}
\end{figure}

Now, let us consider the impedance from one lead to the nanotube.
It should be obvious by now that the capacitive contact cannot be
treated as a lumped capacitance. Rather, the capacitance between the
lead and the nanotube is distributed along the length of the tube. We
must also keep in mind that there is a distributed kinetic inductance
along the length of the tube. This may seem like a difficult problem,
but in fact we have already developed the mathematical machinery
necessary to fully solve this problem.  The impedance from the
macroscopic lead to the nanotube is equal to the impedance from the
nanotube to the lead.  Above, we calculated the impedance from a
nanotube to ``ground''. In the case we are considering here we can use
the results of those calculations, only now instead of the nanotube
coupled to a ground plane, it is coupled to a lead.
Thus, the impedance of the capacitive
coupling to the nanotube is exactly equal to the impedance calculated
in equation~\ref{eq:zinputgeneral}, with $Z_{L}$ equal to infinity. 
Therefore, the impedance from one lead to another is equal to twice
the impedance of equation~\ref{eq:zinputgeneral}. We calculate this numerically
and plot the result in figure~\ref{fig:znanotubefigure6}, for a tube
length of 100~$\mu m$ under each lead, and a very short length of
nanotube between the leads. We use a resistance per length of
100~$\Omega/\mu m$. The resonant behavior is again clear. This
technique may be conceptually the most difficult to understand, but is
in practice the simplest to implement experimentally.

\subsection{Quantum electric field effects}

In the above calculations the electromagnetic field is considered
classical. However, at low (and even room\cite{postma}) 
temperatures the capacitive charging energy
can be considered quantized since $e^2/2C$ can be much less than
$k_B T$. Additionally, the electromagnetic field must be considered quantum
mechanically (as photons) if the photon energy $h\nu$ is greater than the
charging energy. This occurs as a typical energy of 0.5~K for a
10~GHz photon. Therefore, if the discreteness of the photon field is
taken into account, a more sophisticated quantum treatment of the
nanotube dynamical impedance, which takes into account processes such
as photon assisted tunneling, will be necessary. Such a treatment is
beyond the scope of this manuscript.

\section{Conclusions}

We have considered the dynamical properties of single-walled carbon
nanotubes from a circuit point of view.  The 1d plasmon should be
observable using the same experimental technique we developed for
measurements of the 2d plasmons. This measurement would be direct
confirmation of Luttinger liquid behavior of a 1d system of interacting
quantum particles. 
We have formulated our experiment in the frequency domain, but it
should also be possible to perform a time domain experiment using
similar principles to measure the wave velocity and damping.

\section{Acknowledgments}
This work was supported by an ONR Young Investigator award
 (N00014-02-1-0456). We thank J.P. Eisenstein and Cees Dekker 
for useful discussions.

%\begin{thebibliography}{10}

%\end{thebibliography}

%\bibliography{tubepaper}

\end{document}